\documentstyle[12pt]{article}

\begin{document}

\begin{titlepage}

\vskip .6in

\begin{center}
{\Large {\bf   Gauging of Nonlinearly Realized Symmetries   }}\\[10pt]

\end{center}

\normalsize
\vskip .6in

\begin{center}

I. N. McArthur
\par \vskip .1in \noindent
 {\it Department of Physics, The University of Western Australia}\\
{\it Nedlands, Australia.  6907}

\end{center}
\vskip 3cm

\begin{center}
{\large {\bf ABSTRACT}}\\
\end{center}

A representation of a subgroup H of a finite-dimensional group G can be used
to induce a nonlinear realization of G. If the nonlinearly realized symmetry
is gauged, then the BRST charge can be related by a similarity
transformation to the BRST charge for the gauged linear realization of H
(plus a cohomologically trivial piece). It is shown that the relation
between the two BRST charges is a reflection of the fact that they can
be interpreted geometrically as expressions for the exterior derivative
on G relative to two different bases, and an explicit expression for the
generator of the similarity transformation is obtained.
This result is applied in an infinite-dimensional setting, where it
yields the similarity transformation used by Ishikawa and Kato to prove
the equivalence of the Berkovits-Vafa superstring with the underlying
bosonic string theory.

\vspace{9cm}
\noindent

\end{titlepage}

\noindent
\section{Introduction}

Recently, nonlinear realizations of superconformal algebras have appeared in
string theory in two separate guises. Gato-Rivera and Semikhatov \cite{gat}
and Bershadsky et al \cite{bersh} have shown that the field content of
noncritical string theories is generally adequate to admit
 a nonlinearly realized twisted  N=2 superconformal symmetry. The
BRST current (or an improved version of it) is one of the odd generators of
the superconformal algebra, so that the nonlinearly realized symmetry encodes
the BRST structure of the theory in some way. The other guise is in the
Berkovits-Vafa construction \cite{BV} in which N=0 bosonic string theories
can be considered
as N=1 superstring theories with a special choice of background. Central to
this construction is an embedding of the Virasoro algebra into a nonlinearly
realized N=1 superconformal algebra. Both of these manifestations of
nonlinearly realized superconformal symmetries can be found in  higher N
superstrings and in W-strings  \cite{bersh,BO,BFW,KST}.

Amplitudes computed for the Berkovits-Vafa superstring are equivalent to
those for the underlying bosonic string \cite{BV}, which is a reflection of
the fact that the BRST cohomologies of the two theories are identical
\cite{BRS}.
A heuristic explanation for this has been advanced by Polchinski \cite{Pol}.
The basic tenet is that one can promote a linear realization
of a symmetry based on a group H to  a nonlinearly realized
symmetry based on a larger group G by the introduction of extra fields
(``Goldstone bosons'' parameterizing G/H). Gauging this nonlinearly realized
symmetry (without kinetic terms for the gauge fields) gives a theory
equivalent to the original one with the linearly realized symmetry gauged,
in that the gauge degrees of freedom associated with G/H precisely cancel
the extra degrees of freedom introduced in extending the symmetry. In the
case of the Berkovits-Vafa embeddings, the fermionic primary fields
$(b_1, c_1)$ of conformal weight $(\frac32, -\frac12)$ which are introduced
to extend a linearly realized critical Virasoro symmetry  to a nonlinearly
realized critical N=1 superconformal algebra are effectively cancelled
by the bosonic ghosts $(\beta, \gamma)$ of the superstring, leaving a
critical bosonic string theory.

For a related problem (the embedding of a Kac-Moody algebra into an N=1
super Kac-Moody algebra), Figueroa-O'Farrill has carried out an
analysis of the above mechanism at a cohomological level \cite{Fig}. A simpler
approach
is adopted by Ishikawa and Kato \cite{IK} who demonstrate by explicit
construction that there is a similarity transformation which maps
the BRST charge $Q_{N=1}$ of the Berkovits-Vafa superstring to that of the
bosonic string $Q_{N=0}$ (plus a cohomologically trivial topological charge
$Q_{TOP}$)\footnote{This construction has been extended to the case
of the embedding of N=1 superstrings as special backgrounds of N=2
superstrings in \cite{OP}, and to hierarchies of w strings and
super w strings in \cite{KST}.},
\begin{equation}
e^R \, Q_{N=1} \, e^{-R} = Q_{N=0} + Q_{TOP}.
\label{sim}
\end{equation}
The existence of this transformation guarantees the equivalence of the BRST
cohomologies of the two theories \cite{Fig}. Although Ishikawa and Kato
state the form of the charge R in (\ref{sim}), they provide no general means
for determining it. The homological methods of \cite{Fig} are claimed to be
generally applicable, but given the simplicity of the approach adopted
by Ishikawa and Kato, it is important to obtain a better understanding of the
nature of the charge R in cases more general than that treated in \cite{IK}.

The aim of this paper is to set in place a formalism which allows the
calculation of the charge R in similarity transformations of the type
(\ref{sim}) relating the BRST charge of a gauged nonlinear extension of a
symmetry to that for the underlying linearly realized symmetry. The problem
is tackled only at the level of nonlinear realizations of finite dimensional
symmetry algebras (as opposed to infinite dimensional symmetry algebras
such as Kac-Moody algebras and superconformal algebras). However, it is
shown that at least in the case of the Berkovits-Vafa construction, the
results derived in the finite dimensional setting can be applied in the
infinite dimensional setting without ``quantum correction.'' It would be
interesting to know whether this is a general phenomenon.

This absence of quantum corrections to R is to be contrasted with the
situation for the nonlinearly realized superconformal algebra itself, where
only part of the structure of this infinite dimensional algebra can be
deduced by applying the standard formalism for nonlinear realizations
of finite dimensional algebras \cite{Kun,McA}. The quantum corrections
have to be determined ``by hand'' using the requirement that the nonlinearly
realized generators close under operator product expansion. In fact,
(\ref{sim}) provides a systematic method for deriving the quantum corrections
to the nonlinear realization -- knowledge of R and the form of $Q_{N=0}$
and $Q_{TOP}$ allows $Q_{N=1}$ to be determined, from which the nonlinearly
realized generators  can be read. These generators must include the quantum
corrections as the construction (\ref{sim}) ensures that $Q_{N=1}^2$ = 0.
Again, if
R can be determined fully in other situations by application of the
finite dimensional results derived in this paper, then quantum corrections
to nonlinearly realized symmetries could be obtained by similar means.
An example is the twisted N=2 superconformal algebra underlying noncritical
string theories. The ``classical'' form of this algebra can be found
by Hamiltonian reduction \cite{bersh}, but it suffers quantum corrections.
These can be determined by applying quantum Hamiltonian reduction \cite{ham},
which is a fairly complicated procedure (although recent advances \cite{deB}
allow a systematic approach).
If the results of this paper can be applied, considerable simplification
may follow.

The content of the paper is as follows.
In \S2, formalism relating to induced representations of finite-dimensional
groups is introduced,  and expressions for the nonlinearly realized
generators of a group G induced from a linear realization of a subgroup H
are obtained. The BRST charge $Q_{G}$ for the gauged nonlinearly realized
symmetry is constructed in \S3, and it is shown that there exists a charge
R bilinear in the ghost fields such that
\begin{equation}
 e^R \, Q_{G} \, e^{-R} = Q_H + Q_0,
\label{simf}
\end{equation}
where $Q_H$ is the BRST charge associated with the gauged linearly realized
symmetries and $Q_0$ is a cohomologically trivial charge. {\em In particular,
it will be seen that the nilpotent charges $Q_G$ and $Q_H + Q_0$ can
be interpreted geometrically as
representations  of the exterior derivative on
G relative to different bases}.  The expression
obtained for R in the finite-dimensional case is applied in \S4 to the
infinite-dimensional case of the Berkovits-Vafa embedding, where it is
shown to yield the charge R in (\ref{sim}) found by Ishikawa and Kato
\cite{IK}.

It should be stressed that the work in this paper simply aims to obtain an
algorithm which allows the charge R in  (\ref{simf}) to be computed, and
 the mathematical formalism adopted is tailored to this requirement. It is
quite probable that these results can be formulated more elegantly in more
sophisticated mathematical language.

\noindent
\section{Induced Representations}

A representation (or linear realization) of a group H can be used to induce
a nonlinear realization of a group G containing H as a subgroup. At a more
mathematical level, the induced representation is on sections of a
vector bundle over G/H associated to G, which can be considered as a
principal H-bundle over G/H (see, for example, \cite{W}). Induced
representations occur widely in physics. An example is spontaneous symmetry
breaking in quantum field theory, where the unbroken generators of the
symmetry are realized
nonlinearly and the broken generators are realized linearly on the Goldstone
modes associated with the spontaneous symmetry breaking \cite{CWZ,CCWZ}.

Initially we will restrict attention to finite-dimensional groups. Let
$T_a$ $(a =$ $ 1, 2, \cdots , {\rm dim \,G})$ denote the generators of G,
chosen
so that    $T_i$ $(i = 1, 2, \cdots ,$ $ {\rm dim\, H})$ are the generators
of the subgroup H, with the remaining generators $X_{\alpha}$
$(\alpha = 1, 2, \cdots , {\rm dim \,G} - {\rm dim\, H})$ spanning the
complement of {\bf {\cal h}} in {\bf {\cal g}} (where
{\bf {\cal h}} and {\bf {\cal g}} are the
Lie algebras of H and G respectively). Assuming that the coset G/H is
symmetric, then the structure constants $f_{ab}^{\,\,\,\,c}$ of {\cal g} are
determined by
\begin{eqnarray}
\, [T_i , T_j] &=& f_{ij}^{\, \,\,\, k} T_k      \nonumber \\
\, [T_i , T_{\alpha}] &=& f_{i\alpha}^{\, \,\,\,\, \beta} T_{\beta}
\nonumber \\
\, [T_{\alpha} , T_{\beta}] &=& f_{\alpha \beta}^{\, \,\,\,\,\, k} T_k
\label{comm}
\end{eqnarray}
(summation over repeated indices).
Elements of G can be parameterized (at least in a neighbourhood of the
identity) in the form\footnote{ We adopt the (unconventional) right coset
parameterization
to obtain a nonlinear realization which acts by right derivatives of
$\xi^{\alpha}.$}
\begin{equation}
 g(y, \xi) = e^{y^i T_i} \, e^{\xi^{\alpha} X_{\alpha}}.
\label{param}
\end{equation}
The coordinates $\xi^{\alpha}$ parameterize a slice through the identity
locally
isomorphic to H/G, and the $y^i$ parameterize the H-orbits through points
on this slice.

The Lie algebra valued one-form $g^{-1}dg$ on G can be
decomposed as
\begin{equation}
g^{-1}dg = \omega^a T_a,
\label{Maurer}
\end{equation}
where the $\omega^a$ form a basis of left-invariant one-forms on G.
The Maurer-Cartan equation is
\begin{equation}
d \omega^a = -\frac12 f_{bc}^{\,\,\,\, a} \, \omega^b \wedge \omega^c.
\label{MaurerCartan}
\end{equation}
Introducing one-forms $\phi^i$ on the H-orbits by
$$ e^{-y^i T_i}d \,e^{y^i T_i} = \phi^i(y) T_i,$$
then  (\ref{param}) yields
\begin{eqnarray*}
g^{-1}dg &=& e^{-\xi^{\alpha}X_{\alpha}} \, \phi^i T_i
\,e^{\xi^{\alpha}X_{\alpha}} +
e^{-\xi^{\alpha}X_{\alpha}}d\,\,e^{\xi^{\alpha}X_{\alpha}} \\
&=& \phi^i (T_i - A_i^{\,\, \alpha}X_{ \alpha} + \frac{1}{2!}
A_i^{\,\,\alpha} A_{\alpha}^{\,\,\,\, j} T_j - \cdots) \\
&& + d\xi^{\alpha} ( X_{\alpha} - \frac{1}{2!} A_{\alpha}^{\,\,\,\, i} T_i
+ \frac{1}{3!}A_{\alpha}^{\,\,\,\, i}A_i^{\,\, \beta}X_{ \beta} - \cdots)
\end{eqnarray*}
where
\begin{equation}
A_{\alpha}^{\,\,\,\, i} = \xi^{\beta} f_{ \beta \alpha }^{\,\,\,\,\,\,\,i},
\,\,\,\,\,\,\,\,\,\,\,\,
A_i^{\,\,\alpha} = \xi^{\beta} f_{\beta i}^{\,\,\,\,\, \alpha}.
\label{A}
\end{equation}
Thus it follows from (\ref{Maurer}) that
\begin{equation}
\omega^i = \phi^j \, ( {\rm 1} + \frac{1}{2!} A^2 + \frac{1}{4!} A^4 +
\cdots )_j^{\,\,\, i}
 - d\xi^{\beta} \, (\frac{1}{2!} A + \frac{1}{4!} A^3 +
\cdots)_{\beta}^{\,\,\,\,i}
\label{wi}
\end{equation}
and
\begin{equation}
\omega^{\alpha} = -\phi^j \, ( A + \frac{1}{3!} A^3 +
\cdots )_j^{\,\, \, \alpha}
+ d\xi^{\beta} \, ( {\rm 1} + \frac{1}{3!} A^2 + \frac{1}{5!} A^4 +
\cdots)_{\beta}^{\,\,\,\, \alpha}.
\label{walpha}
\end{equation}

Let $Y_a$ denote a basis of left-invariant vector fields on G dual to the
basis $\omega^a$ of left-invariant one-forms, $\omega^a (Y_b) =
\delta^a_{\,\,b},$ and let $\eta_i$ be vector fields on the H-orbits dual
to the one-forms $\phi^i,$ $\phi^i (\eta_j) = \delta^i_{\,\, j}.$
Since
\begin{equation}
d\phi^i = -\frac12 f_{jk}^{\,\,\,\,\, i} \phi^j \wedge \phi^k,
\label{dphi}
\end{equation}
 the vector fields $\eta_i$ provide a realization of
the subalgebra H of G,
\begin{equation}
[\eta_i, \eta_j] = f_{ij}^{\,\,\,\,k} \eta_k.
\label{eta}
\end{equation}
Similarly the Maurer-Cartan equation (\ref{MaurerCartan}) implies
\begin{equation}
[Y_a, Y_b] = f_{ab}^{\,\,\,\, c} Y_c.
\label{Ycomm}
\end{equation}
The vector fields $Y_a$ can be decomposed relative to the basis
$(\eta_i, \frac{\partial}{\partial \xi^{\alpha}})$ of vector fields,
$$ Y_a = Y_a^{\,\, i} \eta_i + Y_a^{\,\, \alpha} \frac{\partial }
{\partial \xi^{\alpha}}.$$
It follows from (\ref{wi}), (\ref{walpha}) and $\omega^a(Y_b) =
\delta^a_{\,\, b}$ that\footnote{A more compact notation could
be introduced, for example $Y_{\alpha}^{\,\,\, \beta} = \bigl[ A (\sinh A)^{-1}
\cosh A\bigr]_{\alpha}^{\,\,\,\, \beta},$ $Y_{\alpha}^{\,\,\, j} =
\bigl[ (\sinh A)^{-1} (\cosh A - 1)]_{\alpha}^{\,\,\,\, j}.$}
\begin{eqnarray}
Y_i^{\,\, j} &=& \delta_i^{\,\, j} \nonumber \\
Y_i^{\,\, \beta} &=& A_i^{\,\,\beta} \nonumber \\
Y_{\alpha}^{\,\,\, \beta} &=& \biggl[ ( {\rm 1} +   \frac{1}{3!} A^2 +
 \frac{1}{5!} A^4 + \cdots)^{-1}\, ({\rm 1} + \frac{1}{2!} A^2 +
\frac{1}{4!} A^4 + \cdots)\biggr]_{\alpha}^{\,\,\, \beta} \nonumber \\
Y_{\alpha}^{\,\,\, j} &=&  \biggl[ ( {\rm 1} +   \frac{1}{3!} A^2 +
 \frac{1}{5!} A^4 + \cdots)^{-1}\, (\frac{1}{2!} A + \frac{1}{4!} A^3 +
\cdots) \biggr]_{\alpha}^{\,\,\,\, j}.
\label{Y}
\end{eqnarray}
Using these results,
\begin{eqnarray}
Y_i &=& \eta_i + A_i^{\,\,\alpha} \frac{\partial}{\partial \xi^{\alpha}}
\nonumber \\
Y_{\alpha} &=& \frac{\partial}{\partial \xi^{\alpha}} + \frac12
A_{\alpha}^{\,\,\,\,i} \eta_i + \frac13 A_{\alpha}^{\,\,\,\,i}
A_i^{\,\,\beta} \frac{\partial}{\partial \xi^{\beta}}
- \frac{1}{24} A_{\alpha}^{\,\,\,\,i} A_i^{\,\,\beta}
A_{\beta}^{\,\,\,\,j}\eta_j + O(A^4),
\label{nonlin}
\end{eqnarray}
where higher order terms can be computed using (\ref{Y}).
Since the left-invariant vector fields $Y_a$ satisfy the commutation
relations (\ref{Ycomm}), it then follows that (\ref{nonlin}) is
a nonlinear realization of the Lie algeba of G. If the $\eta_i,$
which satisfy (\ref{eta}), are replaced by a representation of the
generators of H on some vector space with coordinates $v,$
then (\ref{nonlin}) provides a nonlinear realization of G on the space
with coordinates $(v,\xi).$ As expected for a realization  induced
from a representation of H, the subgroup H is realized linearly on this
space (the $A$ are linear in $\xi$).

\noindent
\section{Gauging the Induced Representation}

Gauging the induced representation of G leads to the introduction of
a pair of fermionic ghosts $(b_a, c^a)$  for each
generator of G. These satisfy the anticommutation relation $\{b_a, c^b\}
= \delta_a^{\,\,b}.$ The corresponding BRST charge is
\begin{equation}
Q_G = c^a Y_a + \frac12 c^a T_a^{(gh)},
\label{Q1}
\end{equation}
 where the ghost charges
\begin{equation}
T_a^{(gh)} = - f_{ab}^{\,\,\,\, c} c^b b_c
\label{Tgh}
\end{equation}
satisfy
$[T_a^{(gh)}, T_b^{(gh)}] = f_{ab}^{\,\,\,\, c}\, T_c^{(gh)}.$ By construction
 the BRST charge $Q_G$ is nilpotent, $\{Q_G, Q_G \} = 0.$

As is well known, the BRST charge $Q_G$ can be interpreted geometrically
as an exterior derivative on the manifold G. Relative to the basis
$\omega^a$ if left-invariant one-forms on G, the exterior derivative on G
is
\begin{equation}
d = \omega^a Y_a.
\label{d1}
\end{equation}
In $Q_G$, the role of the one-forms $\omega^a$ is played by the ghost
fields $c^a,$ with $c^a\,Y_a$ reproducing the exterior derivative of
functions on G. The term $ \frac12 c^a\, T_a^{(gh)}$ in $Q_G$ reproduces
the action of the exterior derivative on forms, as determined by
(\ref{MaurerCartan}).

Relative to the basis $(\phi^i, d\xi^{\alpha})$ for one-forms on G, the
exterior derivative takes the form
\begin{equation}
 d = \phi^i \eta_i + d\xi^{\alpha} \frac{\partial}{\partial \xi^{\alpha}}.
\label{d2}
\end{equation}
A corresponding nilpotent charge ${\tilde Q}_G$ can be realized in terms of
fermionic
ghosts $({\tilde b}_a, {\tilde c}^a)$ (with $\{ {\tilde b}_a, {\tilde c}^b\} =
\delta_a^{\,\,\, b}$) by
\begin{equation}
 {\tilde Q}_G = {\tilde c}^i \eta_i - \frac12 {\tilde c}^i (f_{ij}^{\,\,\,\, k}
{\tilde c}^j {\tilde b}_k) + {\tilde c}^{\alpha}
\frac{\partial}{\partial \xi^{\alpha}}.
\label{Q2}
\end{equation}
With the identification $({\tilde c}^i,{\tilde c}^{\alpha})
\rightarrow (\phi^i, d\xi^{\alpha}),$ the terms linear in ${\tilde c}^a$
reproduce the action of the exterior derivative on functions on G. The term
trilinear in the ghosts reproduces the action of the exterior derivative
on $\phi^i,$ as given by (\ref{dphi}). The absence of any trilinear terms
involving the ghost fields $({\tilde b}_{\alpha},{\tilde c}^{\alpha})$ is
a reflection of the fact that $d(d\xi^{\alpha}) = 0.$

The two expressions (\ref{d1}) and (\ref{d2}) for the exterior derivative on
G are simply expressions relative to different bases for vector fields on
G (and the corresponding dual bases of one-forms), namely
\begin{eqnarray*}
d&=& \omega^i Y_i + \omega^{\alpha} Y_{\alpha} \\
&=& \omega^a Y_a^{\,\, i}  \,\eta_i
+ \omega^a Y_a^{\,\, \alpha}
\frac{\partial}{\partial \xi^{\alpha}}.
\end{eqnarray*}
On the other hand,  the ghost fields  playing the role of one-forms
in the the nilpotent charges (\ref{Q1}) and (\ref{Q2}) corresponding to
(\ref{d1}) and (\ref{d2}) respectively are ``passive'' in nature,
in that the action of the exterior
derivative on forms has to be reproduced by terms trilinear in the ghosts.
Consequently, the change of basis which relates $Q_G$ and ${\tilde Q}_G$
must  be
implemented ``actively'' by a transformation of the ghost fields. If
we identify $c^a$ and ${\tilde c}^a,$ then the terms in $Q_G$ and
${\tilde Q}_G$
not involving ghost generators are $c^a Y_a$ $ = c^a Y_a^{\, \, i} \eta_i
+ c^a Y_a^{\,\, \alpha} \frac{\partial}{\partial \xi^{\alpha}}$ and
$c^i \eta_i + c^{\alpha} \frac{\partial}{\partial \xi^{\alpha}}$
respectively. The ``change of basis'' involved in going from the first
expression to the second is the transformation $c^a \rightarrow c^b
(Y^{-1})_b^{\,\, a}.$ Since this is a linear transformation on the vector space
of ghosts $c^a,$ it can be achieved by a similarity transformation generated
by a bilinear R in the $b_a$ and $c^a,$
\begin{equation}
 e^R c^a e^{-R} = c^b (Y^{-1})_b^{\,\, a}.
\label{transf}
\end{equation}
As will be demonstrated later, the  similarity transformation also achieves
the desired change of basis for the terms trilinear in ghost fields,
so that
$$e^R Q_G e^{-R} = {\tilde Q}_G.$$
This is precisely the relation (\ref{simf}) described in the introduction,
with $Q_H = c^i \eta_i - \frac12 c^i (f_{ij}^{\,\,\,\, k}
c^j b_k)$ and $Q_0 = c^{\alpha}
\frac{\partial}{\partial \xi^{\alpha}}.$

To establish the above results, we first determine the form of the
generator R of the transformation. Expressing R in the form
\begin{equation}
R = c^a K_a^{\,\, b} \, b_b,
\label{R}
\end{equation}
then the anticommutation relation $\{b_a, c^b \} = \delta_a^{\,\, b}$ yields
$[R, c^a] = c^b K_b^{\,\,a},$ so that
$$ e^R c^a e^{-R} = c^b (e^K)_b^{\,\, a}.$$
Thus we require $(e^K)_b^{\,\, a} = (Y^{-1})_b^{\,\, a}.$ Letting
$$Y^{-1} = {\rm 1} + Z,$$
then the matrix K is expressed in terms of the matrix Z by
\begin{equation}
K = \ln ({\rm 1} + Z) = Z - \frac12 Z^2 + \frac13 Z^3 -\frac14 Z^4 +
 \cdots \, .
\label{K}
\end{equation}
The matrices $Y^{-1} = {\rm 1} + Z$ can be read from the expressions
(\ref{wi}) and (\ref{walpha}) using $\omega^a = \phi^i (Y^{-1})_i^{\,\, a}
+ d\xi^{\alpha} (Y^{-1})_{\alpha}^{\,\,\, a}.$ The results are
\begin{eqnarray}
Z_i^{\,\, j} &=& (\frac{1}{2!} A^2 +  \frac{1}{4!} A^4 + \cdots )_i^{\,\, j}
= \bigl[ \cosh A - 1 \bigr]_i^{\,\, j}\nonumber \\
Z_{\alpha}^{\,\,\, j} &=& -(\frac{1}{2!} A +  \frac{1}{4!} A^3
 + \cdots )_{\alpha}^{\,\,\, j} = \bigl[ A^{-1} (1 - \cosh A)
\bigr]_{\alpha}^{\,\,\, j} \nonumber \\
Z_i^{\,\, \beta} &=& -( A +  \frac{1}{3!} A^3 + \cdots )_i^{\,\,  \beta}
= - \bigl[ \sinh A \bigr]_i^{\,\, \beta} \\
Z_{\alpha}^{\,\,\,\beta} &=& (\frac{1}{3!} A^2 +  \frac{1}{5!} A^4
+  \cdots )_{\alpha}^{\,\,\, \beta} = \bigl[ A^{-1} \sinh A - 1
\bigr]_{\alpha}^{\,\,\,\beta}. \nonumber
\end{eqnarray}
These expressions allow K and hence R to be determined to any desired
order in $\xi^{\alpha}$ using (\ref{K}) and (\ref{R}). To $O(\xi^3),$
\begin{eqnarray}
R &=& - c^i A_i^{\,\, \alpha} b_{\alpha} - \frac12 c^{\alpha}
 A_{\alpha}^{\,\,\,\, i} b_i  + \frac14  c^i A_i^{\,\, \alpha}
 A_{\alpha}^{\,\,\,\, j} b_j \nonumber \\
& & -\frac12 c^{\alpha} A_{\alpha}^{\,\,\,\, i} A_i^{\,\, \beta} b_{\beta}
+ \frac{1}{24}  c^{\alpha} A_{\alpha}^{\,\,\,\, i} A_i^{\,\, \beta}
A_{\beta}^{\,\,\,\, j} b_j + O(\xi^4),
\label{R3}
\end{eqnarray}
where $A_i^{\,\, \alpha}$ and $A_{\alpha}^{\,\,\,\, i}$ are given by (\ref{A}).

So far, the charge R has been constructed to ensure
$$ (e^R c^a e^{-R}) Y_a = c^i \eta_i + c^{\alpha}
\frac{\partial}{\partial \xi^{\alpha}}.$$
To prove that $e^R\, Q_G e^{-R} = {\tilde Q}_G,$ it is necessary to show that
$$
(e^R c^a e^{-R}) (e^R\, Y_a e^{-R} - Y_a) +\frac12 e^R ( c^a\, T_a^{(gh)})
 e^{-R} = -\frac12 c^i f_{ij}^{\,\,\,\, k} c^j b_k.
$$
Note that $e^R \,Y_a e^{-R}$ is nonzero as R is $\xi$ dependent and the
nonlinearly realized generators (\ref{nonlin}) contain $\xi$ derivatives.

Using $ e^R \,c^a e^{-R} = c^b (Y^{-1})_b^{\,\, a}$ and $ e^R\, b_a e^{-R} =
 Y_a^{\,\, b}\, b_b$ and the definition (\ref{Tgh}) of $T_a^{(gh)},$
one obtains
$$ \frac12 e^R ( c^a \,T_a^{(gh)}) e^{-R} = -\frac12 c^d c^e b_f
(Y^{-1})_d^{\,\,\, a} f_{ab}^{\,\,\,\,\, c} \,(Y^{-1})_e^{\,\,\, b}\,
Y_c^{\,\, f}.$$
If we denote the basis $(\eta_i, \frac{\partial}{\partial \xi^{\alpha}})$
of vector fields by $V_a,$ then the only nonvanishing commutator is
$[V_i, V_j] = f_{ij}^{\,\,\,\, k} V_k.$ Using $Y_a = Y_a^{\,\, b} V_b$
and (\ref{Ycomm}),
\begin{eqnarray*}
 -\frac12 c^d c^e &b_f& (Y^{-1})_d^{\,\,\, a} f_{ab}^{\,\,\,\,\, c} \,
(Y^{-1})_e^{\,\,\, b}\, Y_c^{\,\, f} = -\frac12 c_i f_{ij}^{\,\,\,\, k}
c^j b_k \\
& & + \frac12 c^dc^eb_f\biggl( (Y^{-1})_d^{\,\,\, c} \,(V_e.Y_c^{\,\, f})
- (Y^{-1})_e^{\,\,\, c}\, (V_d. Y_c^{\,\, f}) \biggr),
\end{eqnarray*}
where $V_e.Y_c^{\,\, f}$ denotes the Lie derivative of the function
$Y_c^{\,\, f}$ with respect to the vector field $V_e$ (the ``directional''
derivative).
Thus it remains to show that
\begin{equation}
 c^d c^e b_f  (Y^{-1})_d^{\,\,\, c}\, (V_e.Y_c^{\,\, f})
= -(e^R c^a e^{-R}) (e^R \,Y_a e^{-R} - Y_a).
\label{show}
\end{equation}
Using  $ e^R c^a e^{-R} = c^b (Y^{-1})_b^{\,\, a}$ and $Y_a =
Y_a^{\,\, b} V_b,$ the right hand side of this expression is $-c^a(e^R \,
V_a e^{-R} - V_a).$ Now, since by  (\ref{R}),  $V_a.R = c^b(V_a.K_b^{\,\, c})
b_c,$ and $[c^a B_a^{\,\, b}\, b_b, R] = c^a[B,K]_a^{\,\,\, b} \,b_b$ for
any matrix B, it follows that
\begin{eqnarray*}
e^R\, V_a e^{-R} - V_a &=&  - V_a.R +\frac{1}{2!} [V_a.R, R] - \frac{1}{3!}
[[V_a.R, R],R] +\cdots \\
&=& c^b ( -V_a.K  +\frac{1}{2!} [V_a.K, K] - \frac{1}{3!}
[[V_a.K, K],K] +\cdots)_b^{\,\, c}\, b_c \\
&=& c^b(e^K V_a e^{-K} - V_a)_b^{\,\,c}\, b_c\\
&=& c^b(Y^{-1} V_a Y - V_a)_b^{\,\,c}\, b_c.
\end{eqnarray*}
This yields
$$-(e^R c^a e^{-R}) (e^R\, Y_a e^{-R} - Y_a) = c^b c^a(Y^{-1} V_a Y -
 V_a)_b^{\,\, c}\,b_c$$
which is precisely the required result (\ref{show}).
So we have indeed established that
$$ e^R Q_G e^{-R} = {\tilde Q}_G.$$
This has been explicitly checked by hand to $O(\xi^3)$ using expression
(\ref{R3}) for R.

\noindent
\section{An Infinite Dimensional Application}

Although the results of the previous section were derived in the context of
finite-dimensional groups, it will be shown here that the result (\ref{sim})
of Ishikawa and Kato showing the triviality of the Berkovits-Vafa
embedding is an example of this construction in an infinite-dimensional
context.

As mentioned in the introduction, the essence of the Berkovits-Vafa
construction is that by starting with a c=26 Virasoro algebra (determining
a critical string theory), it is possible to extend this to a nonlinearly
realized critical (c=15) super-Virasoro algebra by the inclusion of  fermionic
primary fields\footnote{These fields were denoted (b,c) in
\cite{McA} where no confusion
with the standard ghosts of string theory could arise. We adopt the more
standard \cite{BV} notation $(b_1, c_1)$ in this paper.}
 $(b_1, c_1)$ of conformal weight $(\frac32, -\frac12).$
Specifically, if T denotes the generator of the c=26 Virasoro algebra,
then the nonlinearly realized c=15 super-Virasoro algebra has generators
${\tilde T}(z)$ and ${\tilde G}(z)$ given by \cite{BV}
\begin{eqnarray}
{\tilde T}(z) &=& T(z) - \frac32 {\bf :} b_1 \partial c_1 {\bf :}(z)
-\frac12 {\bf :}\partial b_1 c_1{\bf :}(z) + \frac12 \partial^2(c_1
\partial c_1)(z)
\nonumber \\
{\tilde G}(z) &=& b_1(z) + c_1 T(z) +{\bf :}b_1 c_1\partial c_1{\bf :}(z)
+ \frac52 \partial^2 c_1(z) .
\label{sV}
\end{eqnarray}

Gauging the nonlinearly realized super-Virasoro algebra (\ref{sV})
results in the introduction of fermionic ghosts $(b,c)$ of conformal
weight $(2,-1)$ and bosonic ghosts $(\beta, \gamma)$ of conformal weight
$(\frac32, -\frac12).$ In terms of the Berkovits-Vafa construction in
which the critical bosonic string is regarded as a critical superstring
in a  special background, these are the standard superconformal ghosts.
The corresponding BRST charge is
\begin{eqnarray}
Q_{N=1} &=& \oint dz ( c{\tilde T} -\frac12 \gamma {\tilde G} -
{\bf :}c{\bf :}b \partial c{\bf :}{\bf :} -\frac14 b \gamma^2 \nonumber \\
& &+\frac12 \partial c{\bf :}\beta \gamma{\bf :} - c {\bf :}\beta \partial
\gamma{\bf :})(z),
\label{sQ}
\end{eqnarray}
and is nilpotent due to the fact that the central charge c=15 of the
super-Virasoro algebra is the critical value.

As shown by explicit construction in \cite{IK}, there exists a charge R
bilinear in ghost fields such that
$$e^R Q_{N=1} e^{-R} = Q_{N=0} + Q_{TOP},$$
where $Q_{N=0} = \oint dz \,(c T - {\bf :}c{\bf :}b\partial c{\bf :}{\bf :})
(z)$ is the BRST charge of the critical Virasoro algebra generated by T
and $Q_{TOP} = -\frac12 \oint dz \,b_1\gamma (z)$ is a BRST charge associated
with a cohomologically trivial topological sector composed of the
fields $(b_1,c_1)$ and $(\beta, \gamma).$ The charge R is given by Ishikawa
and Kato as
\begin{equation}
R= \oint dz c_1 (\frac12 \gamma \beta - 3 \partial c \beta - 2 c \partial
\beta - \frac12 \partial c_1 {\bf :}c b {\bf :} +
\frac14 \partial c_1 {\bf :}  \gamma \beta{\bf :} )(z).
\label{RIK1}
\end{equation}
We will show that this result can be obtained by a naive application
of the results of the previous section.

To extend the results of \S3 to this case, two generalizations must
be made. The first is that we are dealing with a superalgebra, and the
analogues of the generators $X_{\alpha}$ associated with the coset G/H
are odd generators (namely, the modes of the supercurrent). Correspondingly
the analogues of the parameters $\xi^{\alpha}$ are Grassmann-odd and
the analogues of the ghosts $b_{\alpha}$ and $c^{\alpha}$ are bosonic.
Careful analysis of the steps in the the previous section leading to the
expression for R shows that it is unchanged in the case of a
superalgebra {\em provided \/} that the bosonic ghosts $b_{\alpha}$ and
$c^{\alpha}$ satisfy the commutation relation $[b_{\alpha}, c^{\beta}] =
\delta_{\alpha}^{\,\,\,\, \beta}$ (the fermionic ghosts $b_i$ and $c^i$
still satisfy
$\{b_i, c^j\} = \delta_i^{\,\, j}$). This differs in sign from the usual
convention for bosonic ghosts, and must be borne in mind when comparing
with the results of Ishakawa and Kato.

The second extension is from a finite-dimensional superalgebra {\bf {\cal g}}
 and
subalgebra {\bf {\cal h}} to an  infinite dimensional one, namely the
super-Virasoro
algebra with a critical (c=26) Virasoro subalgebra. Thus the even generators
$T_i$ of {\bf {\cal h}} are are replaced by the modes $L_n$ of the Virasoro
algebra (along with a central term which will be denoted ${\bf 1}$),
the role of the odd generators $X_{\alpha}$ of {\bf {\cal g}} is played
 by the modes
$G_m$ of the supercurrent\footnote{We work in the Ramond sector.},
and the  commutation relations
(\ref{comm}) are replaced by
\begin{eqnarray}
\,[L_m, L_n] &=& (m-n) L_{m+n} + \frac{13}{6} (m^3 - m) \, \delta_{m+n,0}\,
{\bf 1} \nonumber \\
\,[L_m, G_n] &=& (\frac{m}{2} - n) G_{m+n} \nonumber \\
\, \{G_m, G_n\} &=& 2 L_{m+n} + \frac{26}{3} (m^2 - \frac14) \,
\delta_{m+n,0} \, {\bf 1}.
\label{scomm}
\end{eqnarray}
The analogues of the nonlinearly realized generators $Y_i$ and $Y_{\alpha}$
of G in (\ref{nonlin})  are the modes
  ${\tilde L}_m$ and ${\tilde G}_m$ of the generators
${\tilde T}(z)$ and ${\tilde G}(z)$ of the nonlinearly realized
 super-Virasoro algebra (\ref{sV}).
As detailed in \cite{McA}, the role of the Grassmann-odd parameter
$\xi^{\alpha}$ and its derivative $\frac{\partial}{\partial \xi^{\alpha}}$
are played
 respectively by the modes $(c_1)_{-m}$ and $(b_1)_m$ of the fermionic
 fields $c_1(z)$ and $b_1(z),$ the anticommutation relation $\{
\frac{\partial}{\partial \xi^{\alpha}}, \xi^{\beta}\} = \delta_{\alpha}^
{\,\,\, \beta}$ being replaced by $\{(b_1)_m, (c_1)_n\} = \delta_{m+n,0}$.
The expressions (\ref{sV}) for the nonlinearly realized generators of the
super-Virasoro algebra {\em cannot \/} be obtained by a simple translation of
the finite-dimensional results (\ref{nonlin}); there are nontrivial quantum
corrections which must be added to the generators to obtain a closed
super-Virasoro algebra (with c=15) \cite{McA}.

To form the BRST charge associated with the nonlinearly
realized algebra (\ref{sV}), the fermionic ghosts $b_i$ and $c^i$ in
(\ref{Q1}) must be replaced by  the modes $b_m$ and $c_{-m}$ of the
superstring ghosts $b(z)$ and $c(z)$ respectively (with $\{ b_m, c_n\}
= \delta_{m+n,0}$). Special care is required in the identification of the
bosonic ghosts $b_{\alpha}$ and $c^{\alpha}$ with the modes of
their infinite-dimensional
analogues $\beta$ and $\gamma.$ The required identification is of
$(b_{\alpha}, c^{\alpha})$ with $(2\beta_m, -\frac12 \gamma_{-m}).$ The sign
is necessary to ensure that the commutation relation
$[b_{\alpha}, c^{\beta}] =  \delta_{\alpha}^{\,\, \beta}$ maps to the
standard bosonic ghost commutation relation $[\beta_m, \gamma_n] =
-\delta_{m+n,0}.$ The factors of 2 are associated with the normalization of
the terms in the BRST current: it contains $c^{\alpha} Y_{\alpha}$ in the
finite-dimensional case (\ref{Q1}), but $-\frac12 \gamma_{-m} {\tilde G}_m$ in
the infinite-dimensional case (\ref{sQ}).

The final ingredients required are the analogues of the matrices
$A_{\alpha}^{\,\,\,\, i} = \xi^{\beta} f_{\beta \alpha}^{\,\,\,\,\,\, i}$ and
$A_i^{\,\, \alpha} = \xi^{\beta} f_{\beta i}^{\,\,\,\,\,\alpha}.$
To avoid confusion of indices in the infinite-dimensional case (where
$i \rightarrow m$ and $\alpha \rightarrow m$), it is convenient to denote
$A_i^{\,\, \alpha} = \xi^{\beta} f_{\beta i}^{\,\,\,\,\,\alpha}$
by $B_i^{\,\, \alpha}.$ Then it follows by comparison of (\ref{comm})
and (\ref{scomm}) that
$$ A_m^{\,\,\, p} = 2 (c_1)_{m-p}, \, \, \, \, \, \, \, \, \, \,
A_m^{\,\,{\bf 1}} = \frac{26}{3} (m^2-\frac14) \, (c_1)_m $$
(where the index 1 is associated with the central term ${\bf 1}$
in the super-Virasoro algebra) and
$$ B_m^{\,\,\, p} = (p-\frac32 m) \, (c_1)_{m-p} , \, \,\,\,\,\,\,\,\,\,
 B_{{\bf 1}}^{\,\, p} = 0.$$

Putting these results together, the direct translation of the expression
(\ref{R3}) to the infinite-dimensional case is
\begin{eqnarray}
R &=& \frac12 \, (c_1)_{-m-p} \gamma_m b_p + 2(p+\frac32)\, (c_1)_{-m-p} c_m
\beta_p \nonumber \\
 & &+\frac12 (-p + \frac{m}{2}) \,(c_1)_p (c_1)_{-m-n-p} c_m b_n
 + \frac14 (p + \frac{n}{3}) \,(c_1)_p (c_1)_{-m-n-p} \gamma_m \beta_n
\nonumber \\
&&+\frac{1}{8}(p-\frac{2n}{3})\, (c_1)_{-m-p} (c_1)_{-n+p}(c_1)_{n-q}
\gamma_m b_q  + O(c_1^4).
\label{Rs}
\end{eqnarray}
In fact, the term cubic in $c_1$ and the $O(c_1^4)$ terms vanish. This
is because
they contain $(ABA)_m^{\,\,\,\, p},$ which can be seen to vanish by
cycling the mode sums involving the $(c_1)_n.$
Using the standard mode decompositions
\begin{eqnarray*}
b_1(z) &=& \sum_n (b_1)_n z^{-n-\frac32}, \, \,
c_1(z) = \sum_n (c_1)_n z^{-n+\frac12}, \, \,
b(z) = \sum_n b_n z^{-n-2},\\
c(z) &=& \sum_n c_n z^{-n+1}, \, \,
\beta(z) = \sum_n \beta_n z^{-n-\frac32}, \, \,
\gamma(z) = \sum_n \gamma_n z^{-n+\frac12},
\end{eqnarray*}
the result (\ref{Rs}) can be put in the form
\begin{equation}
R = \oint dz \, c_1 \bigl( \frac12 \gamma b - 3 \partial c \beta
- 2 c\partial \beta - \frac12 \partial c_1 c b + \frac14 \partial c_1
\gamma \beta \bigr)(z).
\label{RIK2}
\end{equation}
The expression (\ref{RIK1}) can be obtained from the above result
simply by normal ordering the $cb$ and $\gamma \beta$ terms. In particular,
{\em no quantum corrections are necessary to obtain the correct expression
 (\ref{RIK1})  for  R}.

\noindent
\section{Conclusion}

In this paper, a general formula for the charge R in (\ref{simf}) has been
obtained in a finite-dimensional setting. It demonstrates that
if extra degrees of freedom are introduced to induce a nonlinear
realization of G from a representation of a subgroup H, then gauging
the nonlinearly realized symmetry ``undoes'' the construction \cite{Pol}.
A naive extension of the results to the case of the nonlinear realization
of the N=1 superconformal algebra induced from a representation of the
critical Virasoro subalgebra has been shown to reproduce the result
(\ref{sim}) of Ishikawa and Kato \cite{IK}.

As has already been noted, the structure of the nonlinearly realized
super-Virasoro algebra (\ref{sV}) can in part be derived using the standard
theory of nonlinear realizations \cite{Kun,McA} (in fact, the results in
\cite{McA} can be reproduced by applying the  identifications in \S4 to
the formulas (\ref{nonlin})). However, there are quantum corrections which
must be included to make the algebra close. A knowledge of R allows these
to be computed, as (\ref{sim}) can be inverted to yield $Q_{N=1}$:
$$ Q_{N=1} = e^{-R} (Q_{N=0} + Q_{ TOP}) e^R$$
where, from (\ref{Q2}),
$Q_{N=0} = \oint dz (c T - {\bf :}c{\bf :}b\partial c{\bf :}{\bf :})(z)$ and
$Q_{ TOP} = -\frac12 \oint dz \, b_1 \gamma(z).$
The nonlinearly realized generators ${\tilde T}(z)$ and ${\tilde G}(z)$
can then be read from $Q_{N=1}$ as the coefficients of the terms
linear in $c(z)$ and $-\frac12 \gamma(z).$ They are guaranteed to include the
quantum corrections, as only the generators with these present will give
a nilpotent BRST charge $Q_{N=1}$: the nilpotency of $Q_{N=1}$ follows
by (\ref{sim}) from that of $Q_{N=0} + Q_{TOP},$ which in turn follows
because $T(z)$ is the generator of a Virasoro algebra with c=26.

It would be interesting to know whether the finite-dimensional results
of \S3 for R can be applied in general infinite-dimensional settings
without the need for quantum correction. If so, the procedure outlined
in the above paragraph would allow a systematic derivation of ``quantum
nonlinear realizations'' for superconformal algebras with N$>$1. An
example is the nonlinearly realized twisted N=2 superconformal algebra
underlying noncritical string theories \cite{bersh}, which at present
must be obtained by quantum Hamiltonian reduction \cite{bersh,ham}.

\vspace{2cm}
\noindent
{\Large {\bf Acknowledgement}}\\
I wish to thank Dr J. McCarthy for discussions related to this work.

\end{document}